\documentclass[12pt]{article}
\usepackage{amsmath,xcolor}
\input amssym

\DeclareMathOperator{\diag}{diag}
\begin{document}

\def\prg#1{\medskip\noindent{\bf #1}}  \def\ra{\rightarrow}
\def\lra{\leftrightarrow}              \def\Ra{\Rightarrow}
\def\nin{\noindent}                    \def\pd{\partial}
\def\dis{\displaystyle}                \def\inn{\hook}
\def\grl{{GR$_\Lambda$}}               \def\Lra{{\Leftrightarrow}}
\def\cs{{\scriptstyle\rm CS}}          \def\ads3{{\rm AdS$_3$}}
\def\Leff{\hbox{$\mit\L_{\hspace{.6pt}\rm eff}\,$}}
\def\bull{\raise.25ex\hbox{\vrule height.8ex width.8ex}}
\def\ric{{Ric}}                      \def\tric{{(\widetilde{Ric})}}
\def\tmgl{\hbox{TMG$_\Lambda$}}
\def\Lie{{\cal L}\hspace{-.7em}\raise.25ex\hbox{--}\hspace{.2em}}
\def\sS{\hspace{2pt}S\hspace{-0.83em}\diagup}   \def\hd{{^\star}}
\def\dis{\displaystyle}                 \def\ul#1{\underline{#1}}

\def\hook{\hbox{\vrule height0pt width4pt depth0.3pt
\vrule height7pt width0.3pt depth0.3pt
\vrule height0pt width2pt depth0pt}\hspace{0.8pt}}
\def\semidirect{\;{\rlap{$\supset$}\times}\;}
\def\first{\rm (1ST)}       \def\second{\hspace{-1cm}\rm (2ND)}
\def\bm#1{\hbox{{\boldmath $#1$}}}
\def\nb#1{\marginpar{{\large\bf #1}}}

\def\G{\Gamma}        \def\S{\Sigma}        \def\L{{\mit\Lambda}}
\def\D{\Delta}        \def\Th{\Theta}
\def\a{\alpha}        \def\b{\beta}         \def\g{\gamma}
\def\d{\delta}        \def\m{\mu}           \def\n{\nu}
\def\th{\theta}       \def\k{\kappa}        \def\l{\lambda}
\def\vphi{\varphi}    \def\ve{\varepsilon}  \def\p{\pi}
\def\r{\rho}          \def\Om{\Omega}       \def\om{\omega}
\def\s{\sigma}        \def\t{\tau}          \def\eps{\epsilon}
\def\nab{\nabla}      \def\btz{{\rm BTZ}}   \def\heps{\hat\eps}
\def\bu{{\bar u}}     \def\bv{{\bar v}}     \def\bs{{\bar s}}
\def\bx{{\bar x}}     \def\by{{\bar y}}     \def\bom{{\bar\om}}
\def\tphi{{\tilde\vphi}}  \def\tt{{\tilde t}}
\def\tone{(T1)}         \def\ttwo{(T2)}     \def\tB{{\tilde B}}

\def\tG{{\tilde G}}   \def\cF{{\cal F}}      \def\bH{{\bar H}}
\def\cL{{\cal L}}     \def\cM{{\cal M }}     \def\cH{{\cal H}}
\def\hcH{\hat{\cH}}   \def\cE{{\cal E}}      \def\tcE{{\tilde\cE}}
\def\cK{{\cal K}}     \def\hcK{\hat{\cK}}    \def\cT{{\cal T}}
\def\cO{{\cal O}}     \def\hcO{\hat{\cal O}} \def\cV{{\cal V}}
\def\cJ{{\cal J}}     \def\tom{{\tilde\omega}}
\def\cR{{\cal R}}     \def\hR{{\hat R}{}}    \def\hL{{\hat\L}}
\def\tB{{\tilde B}}   \def\tD{{\tilde D}}    \def\tv{{\tilde v}}
\def\tT{{\tilde T}}   \def\tR{{\tilde R}}    \def\tcL{{\tilde\cL}}
\def\hy{{\hat y}\hspace{1pt}}  \def\tcO{{\tilde\cO}}
\def\hC{\hbox{$\hat C$}}

\def\nn{\nonumber}                    \def\vsm{\vspace{-9pt}}
\def\be{\begin{equation}}             \def\ee{\end{equation}}
\def\ba#1{\begin{array}{#1}}          \def\ea{\end{array}}
\def\bea{\begin{eqnarray} }           \def\eea{\end{eqnarray} }
\def\beann{\begin{eqnarray*} }        \def\eeann{\end{eqnarray*} }
\def\beal{\begin{eqalign}}            \def\eeal{\end{eqalign}}
\def\lab#1{\label{eq:#1}}             \def\eq#1{(\ref{eq:#1})}
\def\bsubeq{\begin{subequations}}     \def\esubeq{\end{subequations}}
\def\bitem{\begin{itemize}}           \def\eitem{\end{itemize}}
\renewcommand{\theequation}{\thesection.\arabic{equation}}

\title{Conformally flat black holes in Poincar\'e gauge theory}

\author{M. Blagojevi\'c and B. Cvetkovi\'c\footnote{
        Email addresses: mb@ipb.ac.rs, cbranislav@ipb.ac.rs} \\
Institute of Physics, University of Belgrade \\
                      Pregrevica 118, 11080 Belgrade, Serbia}
\date{\today}
\maketitle

\begin{abstract}
General criteria for the existence of conformally flat Riemannian
solutions in 3D Poincar\'e gauge theory without matter are formulated.
Using these criteria, we show that the Oliva-Tempo-Troncoso black hole, a
solution of the Bergshoeff-Hohm-Town\-send gravity, is also an exact
vacuum solution of the Poincar\'e gauge theory. The related conserved
charges, calculated from the Hamiltonian boundary term, are shown to
satisfy the first law of black hole thermodynamics. The form of the
boundary term is verified by using the covariant Hamiltonian approach.
\end{abstract}
\section{Introduction}
\setcounter{equation}{0}

The use of three-dimensional (3D) gravitational models in the Poincar\'e
gauge theory (PGT), the first properly formulated gauge theory of gravity
\cite{x1,x2,x3,x4}, started in the early 1990s, when Mielke and Baekler
formulated a \emph{topological model} of 3D gravity with torsion
\cite{x5}. Studies of different aspects of the model made a significant
contribution to a proper understanding of the influence of torsion on the
gravitational dynamics; for a recent review, see Blagojevi\'c and Hehl
\cite{x4}, chapter 17. But, as time went on, it eventually became clear
that transition to the level of \emph{quadratic PGT Lagrangians} is
needed, as the existence of propagating torsion modes offers a more
realistic insight into the dynamical role of torsion; for more details,
see Helay\"el-Neto et al. \cite{x6}, Blagojevi\'c and Cvetkovi\'c
\cite{x7}.

It is well known that classical solutions are an important tool for
exploring dynamical content of gravitational theories, including the
quadratic PGT \cite{x4}. Looking at what has been done in 3D, one should
note that the model can accommodate exact torsion waves \cite{x8} and a
Vaidya-like solution with torsion \cite{x9}. Quite interestingly, the
methods used to construct Siklos waves in \cite{x8} are recently
generalized to 4D \cite{x10}.

In order to properly understand the complex dynamical structure of PGT,
powerful Lagrangian and Hamiltonin formalisms have been developed, see
Obukhov \cite{x3}, Chen et al. \cite{x11}, and Ref. \cite{x2,x4}. This
machinery is very useful not only for genuine PGT problems, characterized
by a nonvanishing torsion, but also in studying torsion-free solutions of
PGT. On the other hand, quite recently \cite{x9} we noticed that the issue
of conserved charges of the Oliva--Tempo--Troncoso (OTT) black hole
\cite{x12}, a solution of the Bergshoeff--Hohm--Townsend (BHT) massive
gravity \cite{x13} for the special choice of parameters, is not completely
settled in the literature, see \cite{x14,x15,x16}. Such a situation
motivated us to reconsider the OTT black hole as a Riemannian
(torsion-free) solution of PGT, and try to find the \emph{conserved
charges}, energy and angular momentum, relying on the full power of the
constrained Hamiltonian formalism. The analysis is based on deriving the
\emph{Hamiltonian boundary term}, the values of which correctly reproduce
the conserved charges.

The paper is organized as follows. In section 2, we use the PGT field
equations to study dynamical properties of Riemannian solutions. In
particular, we show that: (i) for a specific condition on the coupling
constants, Riemannian solutions of PGT are conformally flat, and (ii) any
conformally flat solution of the BHT gravity is also a solution of PGT.
The results are used in section 3 to prove that the static OTT black hole
is a solution of PGT. In section 4, we introduce a set of asymptotic
conditions naturally associated to this black hole, and use the
constrained Hamiltonian formalism to construct the improved canonical
generator $\tG$, acting on the related phase space \cite{x17}.
The form of the boundary term in $\tG$ is shown to be directly
related to the OTT asymptotic conditions, and the conserved charges,
defined as the values of $\tG$, are proved to be fully compatible with the
first law of black hole thermodynamics. In section 5, the same approach is
used to analyze the rotating OTT black hole, and in section 6, we
summarize our results and verify the form of the boundary term by
comparing it to the generalized covariant formula proposed by So \cite{x18}.
Appendices contain some technical details.

Our conventions are the same as in Ref. \cite{x9}: the Latin indices $(i,
j, k, ...)$ refer to the local Lorentz frame, the Greek indices
$(\m,\n,\r, ...)$ refer to the coordinate frame, $b^i$ is the orthonormal
triad (coframe 1-form), $\om^{ij}$ is the Lorentz connection (1-form), the
respective field strengths are the torsion $T^i=db^i+\om^i{_m}\wedge b^m$
and the curvature $R^{ij}=d\om^{ij}+\om^i{_k}\wedge \om^{kj}$ (2-forms),
the frame $h_i$ dual to $b^j$ is defined by $h_i\inn b^j=\d_i^j$, the
signature of the metric is $(+,-,-)$, totally antisymmetric symbol
$\ve^{ijk}$ is normalized to $\ve^{012}=1$, the Lie dual of an
antisymmetric form $X^{ij}$ is $X_i:=-\ve_{ijk}X^{jk}/2$, the Hodge dual
of a form $\a$ is $\hd\a$, and the exterior product of forms is implicit.


\section{Conformally flat Riemannian solutions in PGT}\label{sec2}
\setcounter{equation}{0}

The OTT black hole is a vacuum solution of the BHT gravity with a unique
AdS ground state \cite{x12,x14}. Here, based on our earlier experience
\cite{x8,x9}, we wish to interpret it as a Riemannian solution of PGT in
vacuum. By doing so, we will be able to use the full power of the
constrained Hamiltonian formalism to clarify the asymptotic structure
and find the conserved charges for both the static and the rotating OTT
black hole.

The possibility to interpret the OTT black hole as a Riemannian solution
of PGT (a solution with vanishing torsion) is not just a coincidence, it
is based on a deep dynamical relation between the PGT sector of Riemannian
solutions and the BHT gravity. The content of this relation is expressed
by a theorem stating that any conformally flat solution of the BHT gravity
is also a Riemannian solution of PGT. This is, in particular, true for the
OTT black holes. In 3D, the Weyl curvature identically vanishes, and the
Cotton 2-form $C^i$ is used to characterize conformal properties of
spacetime \cite{x19}. It is defined by $C^i:=\nab L^i=dL^i+\om^i{_m}L^m$
where $L^m:=\ric^m-\frac{1}{4}Rb^m$ is the Schouten 1-form. A spacetime is
conformally flat when $C^i=0$.

To prove the above theorem, we note that the BHT gravity action
\be
I_{\rm BHT}=a_0\int d^3x\sqrt{g}\left (R-\l+\frac{1}{m^2}K\right)\,,
\quad K:=\ric^{ij}\ric_{ij}-\frac{3}{8}R^2\, ,                  \nn
\ee
leads to the field equations \cite{x20}:
\bea
&&G_{ij}-\l\eta_{ij}-\frac1{2m^2}K_{ij}=0\, ,                   \lab{2.1}\\
&&K_{ij}=K\eta_{ij}-2L_{ik}G^k{_j}-2(\nab_m C_{in})\ve^{mn}{_j}\,,\nn
\eea
where $G_{ij}=\ric_{ij}-R\eta_{ij}/2$ is the Einstein tensor,
$C_{ij}=h_j\inn\hd C_i$ is the Cotton and $L_{ij}=h_j\inn L_i$ the
Schouten tensor. This compact form of the BHT field equations
significantly simplifies the analysis of conformally flat solutions.

Lagrangian dynamics of PGT is expressed in terms of its basic field
variables, the triad $b^i$ and the Lorentz connection $\om^{ij}$
(1-forms), the related field strengths are the torsion $T^i:=d
b^i+\om^i{_m}b^m$ and the curvature $R^{ij}:=d\om^{ij}+\om^i{_m}\om^{mj}$
(2-forms), and the spacetime continuum is described by a Riemann--Cartan
geometry. The gravitational Lagrangian $L_G=L_G(b^i,T^j,R^{mn})$ (3-form)
is at most quadratic in the field strengths:
\bea
L_G&=&-\hd(a_0R+2\L_0)
    +T^i\,\hd (a_1{}^{(1)}T_i+a_2{}^{(2)}T_i+a_3{}^{(3)}T_i)      \nn\\
 && +\frac{1}{2}R^{ij}\,\hd\left(b_4{}^{(4)}R_{ij}+b_5{}^{(5)}R_{ij}
                       +b_6{}^{(6)}R_{ij}\right)\, ,            \nn
\eea
where ${}^{(n)}T^i$ and ${}^{(n)}R^{ij}$ are irreducible components of the
respective field strengths, and $a_0$ is normalized by  $a_0=/16\pi G$;
for details, see Ref. \cite{x7}. Since we are here interested only in
Riemannian solutions of PGT, the torsion can be effectively set to vanish,
whereas the curvature becomes Riemannian; in 3D, it has only two
nonvanishing irreducible components,
$$
{}^{(6)}R^{ij}=\frac{1}{6}Rb^ib^j\, ,\qquad
{}^{(4)}R_{ij}=R^{ij}-{}^{(6)}R^{ij}\, ,
$$
whereas the third one vanishes, ${}^{(5)}R_{ij}=0$. The Riemannian
reduction of the general field equations takes the form derived in
Appendix A of Ref. \cite{x9}:
\bsubeq\lab{2.2}
\bea
{\rm (1ST)}&&E_i=0\, ,                                          \nn\\
{\rm (2ND)}&&\nab H_{ij}=0\, ,                                  \lab{2.2a}
\eea
where
\bea
&&E_i=h_i\inn L_G-\frac{1}{2}(h_i\inn R^{mn})H_{mn}\, ,         \nn\\
&&H_{ij}=-2a_0\ve_{ijm}b^m +\frac{b_4+2b_6}{6}\,R\ve_{ijk}b^k
         -2b_4\ve_{ij}{^m}L_m\, .                               \lab{2.2b}
\eea
\esubeq

Let us now note a simple property of {\rm (2ND)}: the vanishing of the
second term in $H_{ij}$ implies that the Cotton 2-form $C_m=\nab L_m$
vanishes. More precisely:
\bitem
\item[\tone] A Riemannian solution of PGT is conformally flat iff
    ~$b_4+2b_6=0$.
\eitem
Next, to examine the content of {\rm (1ST)}, it is convenient to express
it in the tensorial form:
$$
a_0\ric_{ij}+2\L_0\eta_{ij}+b_4 L_{im}G^m{_j}=0\, .
$$
In combination with its trace, $a_0R+6\L_0+b_4 K=0$, it can be transformed
to
\be
a_0G_{ij}-\L_0\eta_{ij}
      -b_4\frac{1}{2}\left(K\eta_{ij}-2L_{im}G^m{_j}\right)=0\,.\lab{2.3}
\ee

A direct comparison shows that Eq. \eq{2.3} coincides with the BHT
field equation \eq{2.1} for $C_{in}=0$, provided one makes the following
identification of parameters:
\be
\L_0=a_0\l\,,\qquad b_4={a_0}/{m^2}\, .                         \lab{2.4}
\ee
This leads to the main result of this section:
\bitem
\item[\ttwo] Any conformally flat solution of the BHT gravity is also
    a Riemannian solution of PGT with $b_4+2b_6=0$, and vice versa.
\eitem
An interesting interpretation of the identifications \eq{2.4} is found by
using the BHT condition $\l=-m^2$ that ensures the existence of the
unique maximally symmetric background. For $m^2=1/2\ell^2$, the
identifications \eq{2.4} are transformed into
\be
\L_0=-{a_0}/{2\ell^2}\, ,\qquad b_4=2a_0\ell^2\, .              \lab{2.5}
\ee

Theorems \tone\ and \ttwo\ allow us to study conformally flat
solutions of the BHT massive gravity relying on the powerful Hamiltonan
methods developed in the context of PGT \cite{x2,x4,x11}. In particular,
we will use these methods to study boundary terms, conserved charges, and
central charges of the OTT black hole.
Recently, it was shown by Barnich et al. \cite{y20} that BHT gravity
admits black hole solutions that can be deformed into dynamical ``black
flowers", a new class of solutions that are no longer spherically
symmetric. Since black flowers are conformally flat, they are also
solutions of PGT.

Although PGT is used here as a convenient framework for studying
conformally flat solutions of the BHT gravity, it is worth mentioning some
general dynamical aspects of PGT, expressed through its unitarity properties.
In 3D, the requirement of unitary propagation of torsion modes leads to
certain conditions on the coupling constants, the form of which is given
in Eqs. (17) of Ref. \cite{x6}. The content of these equations leads to
the following conclusions: (a) the condition $b_4+2b_6=0$ implies that the
spin-$0^+$ mode does not propagate; (b) for a suitable choice of the
remaining coupling constants, the propagation of the spin-$0^-$, spin-$1$ or
spin-$2$ modes is unitary.

\section{Static OTT black hole}
\setcounter{equation}{0}

Now, we turn our attention to the static OTT spacetime, described by the
metric \cite{x12}
\be
ds^2=N^2dt^2-\frac{dr^2}{N^2}-r^2 d\vphi^2\, , \qquad
N^2:=-\m+br+\frac{r^2}{\ell^2}\, ,                              \lab{3.1}
\ee
where $\m$ and $b$ are real parameters. The roots of equation $N^2=0$ are
$$
r_\pm=\frac{1}{2}\left(-b\ell^2\pm\ell\sqrt{4\m+{b^2\ell^2}}\,\right)\,.
$$
The OTT metric defines a static AdS black hole when  $\ell^2>0$ and at
least $r_+$ is real and positive; for $b=0$ it reduces to the BTZ black
hole \cite{x21}.

In order to have a suitable geometric description of the OTT black hole in
the framework of PGT, we introduce the triad field (1-form)
\bsubeq\lab{3.2}
\bea
b^0:=Ndt\, ,\qquad b^1:=\frac{dr}{N}\, ,\qquad b^2:=rd\vphi\, , \lab{3.2a}
\eea
so that $ds^2=\eta_{ij}b^i\otimes b^j$, with $\eta=\diag(+1,-1,-1)$, and
the corresponding Riemannian connection (1-form):
\be
\om^{01}=-N'b^0\, , \qquad \om^{02}=0\, ,\qquad
\om^{12}=\frac{N}{r}b^2\, ,                                     \lab{3.2b}
\ee
\esubeq
where $N':=\pd_r N$.
The geometric structure introduced in Eqs. \eq{3.2} can be now used to
calculate first the curvature 2-form $R^{ij}$, and then the Schouten 1-form:
\be
L^0=\frac{1}{2\ell^2}b^0\,,\qquad L^1=\frac{1}{2\ell^2}b^1\, ,\qquad
L^2=\left(\frac{1}{2\ell^2}+\frac{b}{2r}\right)b^2\,.
\ee
An  explicit calculation yields $C^i=\nab L^i=0$, and theorem \ttwo\ from
section \ref{sec2} implies that the static OTT black hole is an exact
Riemannian solution of PGT in vacuum.

It is interesting to compare these general arguments with direct
calculations based on the PGT field equations \eq{2.2}. As shown in
\cite{x9}, the result takes the form of three conditions on the four
Lagrangian parameters $(a_0,b_4,b_6,\L)$:
\be
b_4-2a_0\ell^2=0\,,\qquad a_0+2\ell^2\L_0=0\, ,\qquad
b_4+2b_6=0\, .                                                \lab{3.4}
\ee
The meaning of these conditions is now quite clear: the third one follows
from the conformal flatness of the static OTT black hole, and the first
two coincide with the relations \eq{2.5}.

\section{Asymptotic structure of the static black hole}\label{sec4}
\setcounter{equation}{0}

In this section, we use the canonical approach to analyze the asymptotic
structure naturally associated to the static OTT black hole. In
particular, we wish to calculate the conserved charges and verify their
compatibility with the first law of black hole thermodynamics.

\subsection{Asymptotic conditions}\label{sub41}

The asymptotic state associated to the triad \eq{3.2a} is determined by the
asymptotic formula
$$
N=\frac{r}{\ell}+\frac{b\ell}{2}
  -\frac{\ell}{2r}\left(\m+\frac{b^2\ell^2}{4}\right)+\cO_2\,,
$$
and a similar formula for $1/N$. In order to produce a suitable set of the
asymptotic states, we act on this particular state by the transformations
belonging to the AdS group $SO(2,2)$, as described in Ref. \cite{x7}. The
family of triads obtained in this way has the AdS asymptotic behavior
given by $b^i{_\m}=\bar b^i{_\m}+B^i{_\m}$, where
\be
\bar b^i{_\m}:=\left( \ba{lll}
         \dis\frac{r}{\ell}  & 0  & 0  \\[2pt]
         0 & \dis\frac{\ell}{r}   & 0  \\[7pt]
         0 & 0                    & r
               \ea\right)\, ,
\qquad
B^i{_\m}:=\left( \ba{lll}
         \cO_0  & \cO_3  & \cO_0  \\
         \cO_1  & \cO_2  & \cO_1  \\
         \cO_0  & \cO_3  & \cO_0
               \ea\right)\, .                                   \lab{4.1}
\ee
Here,  $\bar b^i{_\m}$ refers to an AdS background ($b=\mu=0$). Note that
the presence of the OTT parameter $b$ makes the asymptotic decrease of
$B^i{_\m}$ slower then in the BTZ case. The subset of the local Poincar\'e
transformations that respect these conditions is determined by the
parameters $(\xi^\m,\ve^{ij}=-\ve^{ijk}\th_k)$, such that
$$
\d_0 b^i{_\m}
  :=\ve^{ijk}\th_j b_{k\m}-(\pd_\m\xi^\r)b^i{_\r}-\xi^\r\pd_\r b^i{_\m}
   =B^i{_\m}\, .
$$
As a consequence, the asymptotic parameters of local translations and
Lorentz rotations are found to be
\bsubeq\lab{4.2}
\bea
&&\frac{\xi^t}{\ell}=T+\frac{\ell^4}{2r^2}\pd_t^2T+\cO_3\, ,\qquad
  \xi^r=-\ell r\pd_t T+\cO_0\, ,                                    \nn\\
&&\xi^\vphi=S-\frac{\ell^2}{2r^2}\pd_\vphi^2S+\cO_3\, ,             \\
&&\th^0=-\frac{\ell^2}{r}\pd_t\pd_\vphi T+\cO_2\, ,\qquad
  \th^1=\pd_\vphi T+\cO_1\, ,                                       \nn\\
&&\th^2=\frac{\ell^3}{r}\pd_t^2 T+\cO_2\, .
\eea
\esubeq
The functions $T$ and $S$ are such that $\pd_\pm T^\mp=0$, with
$x^\pm:=t/\ell \pm \vphi$, and $T^\pm:=T\pm S$. Thus, in spite of a
relaxed asymptotic behavior of $B^i{_\m}$ as compared to the BTZ black
hole, the values of the corresponding asymptotic parameters are
essentially the same \cite{x22}.

Similar procedure leads to the asymptotic conditions for the connection.
Introducing the Lie dual connection $\om^i$ by $\om^{ij}=-\ve^{ijk}\om_k$,
one finds $\om^i{_\m}=\bar\om^i{_\m}+\Om^i{_\m}$, where
\be
\bom^i{_\m}=\left( \ba{lll}
   0  &  0  &  -\dis\frac{r}{\ell}    \\
   0  &  0  &  0                      \\
   -\dis\frac{r}{\ell^2}  &  0  &  0
           \ea\right)\, ,
\qquad
\Om^i{_\m}:=\left( \ba{lll}
   \cO_0  &  \cO_3 & \cO_0   \\
   \cO_1  &  \cO_2 & \cO_1   \\
   \cO_0  &  \cO_3 & \cO_0
           \ea\right)\, .                                       \lab{4.3}
\ee
The asymptotic behavior of the  connection  does not impose any new
restriction on the asymptotic Poincar\'e parameters \eq{4.2}.

For an easier comparison with the literature, we display here the
deviation of the metric from its background value:
$$
G_{\mu\nu}:=g_{\m\n}-\bar g_{\m\n}
           =\left( \ba{lll}
         \cO_{-1}  & \cO_2  & \cO_{-1}  \\
         \cO_2     & \cO_3  & \cO_2  \\
         \cO_{-1}  & \cO_2  & \cO_{-1}
            \ea\right)\, .
$$

Using the composition law of the asymptotic Poincar\'e parameters \eq{4.2}
to leading order, the commutator algebra of the asymptotic symmetry is
found to have the form of two independent Virasoro algebras,
\be
i[\ell^\pm_m,\ell^\pm_n] = (m - n)\ell^±_{m+n},                 \lab{4.4}
\ee
where $\ell^\pm_n=-\d_0(T^\pm=e^{\pm inx^\pm})$. The respective central
charges $c^\pm$ will be determined by the canonical methods.

The condition $T^i=0$ leads to further asymptotic requirements (Appendix A).

\subsection{Canonical generator and conserved charges}\label{sub42}

The standard construction of the canonical generator for the quadratic PGT
makes use of the existence and classification of all constraints in the
theory. The construction can be significantly simplified by going over to
the first order Lagrangian (3-form)
$$
L_G=T^i\t_i+\frac{1}{2}R^{ij}\r_{ij}-V(b,\t,\r)\, ,
$$
see Refs. \cite{x11,x24}. Here, $\t^m$ and $\r_{ij}$ are independent
dynamical variables, the covariant field momenta conjugate to $b^i$ and
$\om^{ij}$, and the potential $V$ ensures the on-shell relations
$\t_i=T_i$, $\r_{ij}=R_{ij}$, which transform $L_G$ into the standard
quadratic form.

The first order formulation of $L_G$ simplifies the construction of the
canonical generator $G$, the form of which can be found in Ref. \cite{x7},
equation (5.7). Since $G$ acts on the basic dynamical variables via the
Poisson bracket operation, it must be a differentiable functional. To
examine the differentiability of $G$, one starts from the form of its
variation \cite{x8,x9}:
\bsubeq\lab{4.5}
\bea
&&\d G=-\int_\S d^2x(\d G_1+\d G_2)\, ,                         \nn\\
&&\d G_1=\ve^{t\a\b}\xi^\mu\left(b^i{_\mu}\pd_\a\d\t_{i\b}
         +\om^i{_\mu}\pd_\a\d\r_{i\b}+\t^i{_\mu}\pd_\a\d b_{i\b}
         +\r^i{_\mu}\pd_\a\d\om_i{_\b}\right)+\cR\, ,           \nn\\
&&\d G_2=\ve^{t\a\b}\th^i\pd_\a\d\r_{i\b}+\cR\, .
\eea
Here, $\S$ is the spatial section of spacetime, the variation is performed
in the set of adopted asymptotic states, $\cR$ stands for regular
(differentiable) terms, and we use $\r^i$ and $\om^i$, the Lie duals of
$\r_{mn}=H_{mn}$ and $\om_{mn}$, to simplify the formulas.

Using the adopted asymptotic conditions, one finds $\d G_2=\cR$,
which implies
\be
\d G=-\int_\S d^2 x\,\ve^{t\a\b}\xi^\mu\left(
     b^i{_\mu}\pd_\a\d\t_{i\b}+\om^i{_\mu}\pd_\a\d\r_{i\b}
    +\t^i{_\mu}\pd_\a\d b_{i\b}+\r^i{_\mu}\pd_\a\d\om_i{_\b}\right)+\cR.
\ee
\esubeq
Thus, in general, $\d G\ne\cR$ and $G$ is not differentiable. The problem
can be corrected by going over to the improved generator $\tG:=G+\G$,
where the boundary term $\G$ is constructed so that $\d\tG=\cR$ \cite{x17}.
After making a partial integration in $\d G$, one finds that $\G$ is
defined by the variational equation
\be
\d\G=\int_{\pd\S}\xi^\mu\left(
     b^i{_\m}\d\t_i+\om^i{_\m}\d\r_i
    +\t^i{_\m}\d b_i+\r^i{_\m}\d\om_i\right)\, ,                \lab{4.6}
\ee
where $\pd\S$ is the boundary of $\S$ located at infinity, parametrized by
the coordinate $\vphi$. Now, restricting our attention to the Riemannian
sector with $\t^i=0$, we obtain
\bsubeq\lab{4.7}
\be
\d\G=\int_{\pd\S} \xi^\mu\left(
     \om^i{_\m}\d\r_i+\r^i{_\m}\d\om_i\right)
    =\int_0^{2\pi}(\xi^t\d\cE+\xi^\vphi\d\cJ)d\vphi,            \lab{4.7a}
\ee
where (after returning to $\om_{ij}$ and $H_{ij}$)
\bea
&&\d\cE:=\frac{1}{2}\left(\om^{ij}{_t}\d H_{ij\vphi}
                          +H^{ij}{_t}\d\om_{ij\vphi}\right)\, , \lab{4.7b}\\
&&\d\cJ:=\frac{1}{2}\left(\om^{ij}{_\vphi}\d H_{ij\vphi}
                      +H^{ij}{_\vphi}\d\om_{ij\vphi}\right)\,.  \lab{4.7c}
\eea
\esubeq
In what follows, one should take into account that the form
\eq{2.2b} of $H_{mn}$ is simplified after using the restrictions \eq{3.4}
on the Lagrangian parameters:
$$
H_{ij}=-2a_0\ve_{ijk}b^k-4a_0\ell^2\ve_{ijk}L^k\, .
$$

Once we find the solutions for $\cE$ and $\cJ$, the boundary term takes the
form
\be
\G(\xi)=\int_0^{2\pi}(\xi^t\cE+\xi^\vphi\cJ)d\vphi\,.           \lab{4.8}
\ee
In general, Eqs. \eq{4.7} refer to the fields and their variations
belonging to the entire set of asymptotic states, defined by Eqs. \eq{4.1}
and \eq{4.3}. However, it is instructive to consider first a simpler
situation, in which the fields and their variations refer just to a
\emph{single asymptotic state}, the static OTT configuration \eq{3.2}. In
that case, Eq. \eq{4.7b} takes the form
\bea
\d\cE&=&\om^{01}{_t}\d H_{01\vphi}
                          +H^{12}{_t}\d\om_{12\vphi}            \nn\\
 &=&2a_0\ell^2\left(\frac{r}{\ell^2}+\frac{1}{2}b\right)\d b-4a_0N\d N \nn\\
 &=&2a_0\d\left(\m+\frac{1}{4}\ell^2 b^2\right)\, ,             \lab{4.9}
\eea
that is easily integrated to obtain $\cE$. In fact, the procedure just
described is sufficient to calculate the values of the conserved charges,
but only for this particular configuration.

In the next step, we wish to find a solution for $\cE$ on the \emph{whole
set of asymptotic states}. Using the special result \eq{4.9} as a guide,
we find
\bsubeq\lab{4.10}
\bea
&&\cE=\cE_0-\frac{1}{4}\bigl(\D\om^{ij}{_t}\D H_{ij\vphi}
                    +\D H_{ijt}\D\om^{ij}{_\vphi}\bigr)\, ,     \lab{4.10a}\\
&&\cE_0:=\frac{1}{2}\bigl(\om^{ij}{_t} \D H_{ij\vphi}
                    +H_{ijt}\D\om^{ij}{_\vphi}\bigr)\, ,        \nn
\eea
where $\D X:=X-\bar X$ is the difference between any field $X$ and its
boundary value $\bar X$. In a similar manner, Eq. \eq{4.7c} leads to
\bea
\cJ &=&\frac{1}{2}\om^{ij}{_\vphi} H_{ij\vphi}
     =\cJ_0-\frac{1}{2}\D H_{ij\vphi}\D\om^{ij}{_\vphi}\, ,     \lab{4.10b}\\
\cJ_0&:=&\frac{1}{2}\bigl(\om^{ij}{_\vphi} \D H_{ij\vphi}
                       +H_{ij\vphi}\D\om^{ij}{_\vphi}\bigr)\, , \nn
\eea
\esubeq
where the first equality follows directly from \eq{4.7c}, and the second
one from $\bar H_{ij\vphi}\bar\om^{ij}{_\vphi}=0$. With these results for
$\cE$ and $\cJ$, the boundary term \eq{4.8} is seen to to be a finite
phase-space functional that satisfies the variational equation \eq{4.7a}
(Appendix B).

The values of the improved generators for time translations ($\xi=\pd_t$)
and spatial rotations ($\xi=\pd_\vphi$) are given by the corresponding
boundary terms, which define the conserved charges of the system, the
energy and the angular momentum, respectively:
\be
E=\int_0^{2\pi}d\vphi\,\cE\, ,\qquad J=\int_0^{2\pi}d\vphi\,\cJ\, ,
\ee
Calculated on the static OTT configuration, these expressions take
the values
\be
E=\frac{1}{4G}\left(\m+\frac{1}{4}b^2\ell^2\right)\, ,
\qquad J=0\, .                                                  \lab{4.12}
\ee

The expressions \eq{4.10} for $\cE$ and $\cJ$ are obtained by
relying on the set of asymptotic configurations \eq{4.1} and \eq{4.3} that
contain the static OTT black hole geometry. It is interesting to compare
the boundary term \eq{4.8} to the covariant approach of Chen et al.
\cite{x11}. Looking at the Riemannian reduction of their formula (239) and
choosing the upper or lower term in each curly bracket separately, one
finds that none of the resulting expressions can reproduce our result.
To make the argument more clear, consider, for instance, the term
$\cE_0$ in \eq{4.10a} that corresponds to choosing all the upper terms in
(239); the corresponding expression for the energy would be different from
\eq{4.12}: $E_0=\frac{1}{4G}\left(\m+\frac{1}{2}b^2\ell^2\right)$. How do
we know that this result is not correct? The answer can be found by noting
that the boundary term $\G[\xi]$ has a \emph{twofold role:} (i) its values
define the conserved charges, and (ii) its form ensures the improved
generator $\tG=G+\G$ to be a differentiable functional on the phase space
associated with the chosen boundary conditions. Since $\cE_0$ does not
satisfy the variational equation \eq{4.7b}, replacing $\cE$ by $\cE_0$
would destroy the differentiability of the new canonical generator
$\tG[\cE\to\cE_0]$. The way out of this situation can be found in the
work of So \cite{x18}, who proposed a generalized boundary term by
introducing ``mixed" choices involving a linear combinations of upper and
lower term in (239), see footnote ``u" in \cite{x11}. As discussed in
section \ref{sec6}, our boundary term \eq{4.8} is appropriately described
by a particular mixed form. The need for using a mixed boundary term
stems directly from the \emph{slower asymptotic decrease} of the OTT
dynamical variables as compared to the BTZ case (see subsection
\ref{sub41}), or equivalently, from the presence of the $br$ term in the
OTT metric \eq{3.1}.

\subsection{Asymptotic symmetry}\label{sub43}

The results obtained so far allow us to precisely describe the OTT
asymptotic symmetry by the Poisson bracket algebra of the improved
canonical generators. Following the procedure described in \cite{x8,x9},
one finds that this algebra, expressed in terms of the Fourier modes
$L_n^\pm$ of $\tG$, is given by a centrally extended form of the
commutator algebra \eq{4.4},
\be
i[L^\pm_m,L^\pm_n] = (m - n)L^±_{m+n}+\frac{c^\pm}{12}m^3\d_{m,-n},
\ee
where $c^\pm$ are classical central charges,
\be
c^\pm=c\, ,\qquad c=\frac{3\ell}{G}\, .                         \lab{4.14}
\ee

\subsection{Black hole entropy}\label{sub44}

As an additional, theoretical test of the validity of our canonical
expression for the OTT energy \eq{4.12}$_1$, we propose to verify its
exact agreement with the first law of black hole thermodynamics; the
same strategy was used, for instance, by Giribet and Leston \cite{x15}, and
by Maeda \cite{x25}.

The black hole entropy can be calculated from the Cardy formula \cite{x23}
$$
S=2\pi\sqrt{\frac{h^-c^-}6}+2\pi\sqrt{\frac{h^+c^+}6}\,,
$$
where $h^\pm=(\ell E\pm J)/2$. For the static OTT black hole, this formula
yields
\be
S=2\pi\ell\sqrt{\frac{E}{G}}\, .
\ee
Then, using the expression for the Hawking temperature,
\be
T=\frac{1}{4\pi}\left.\pd_r N^2\right|_{r=r_+}
 =\frac{1}{\pi\ell}\sqrt{GE}\, ,
\ee
one can directly verify the first law of the black hole thermodynamics:
\be
\d E=T\d S\, .
\ee
Since the entropy vanishes for $E=0$, the state with $E=0$ can be
naturally regarded as the ground state of the OTT family of black holes
\cite{x14}.

\section{Rotating OTT black hole}
\setcounter{equation}{0}

In order to verify to what extent the canonical expressions \eq{4.10} for
the boundary terms of the static OTT black hole are general, we now use
the same approach to study the rotating OTT black hole.

\subsection{Geometric aspects}

The rotating OTT black hole is defined by the metric \cite{x14,x15}
\bsubeq\lab{5.1}
\be
ds^2=N^2dt^2-F^{-2}dr^2-r^2(d\vphi+N_\vphi dt)^2\,,
\ee
where
\bea
&&F=\frac{H}{r}\sqrt{ \frac{H^2}{\ell^2}
   +\frac{b}{2}H\left(1+\eta\right)
   +\frac{b^2\ell^2}{16}\left(1-\eta\right)^2-\mu\eta }\, ,     \nn\\
&&N=AF\, ,\qquad A=1+\dis\frac{b\ell^2}{4H}(1-\eta)\,,          \nn\\
&&N_{\vphi}=\frac{\ell}{2r^2}\sqrt{1-\eta^2}(\m-bH)\, ,         \nn\\
&&H=\sqrt{ r^2-\frac{\m\ell^2}{2}(1-\eta)
          -\frac{b^2\ell^4}{16}\left(1-\eta\right)^2 }\, .
\eea
\esubeq
The roots of $N=0$ are
$$
r_\pm=\ell\sqrt{\frac{1+\eta}{2}}
    \left(-\frac{b\ell}{2}\sqrt{\eta}
          \pm\sqrt{\m+\frac{b^2\ell^2}{4}}\right)\, .
$$
The metric \eq{5.1} depends on three free parameters, $\m$, $b$ and
$\eta$. For $\eta=1$, it represents the static OTT black hole, and for
$b=0$, it reduces to the rotating BTZ black hole with parameters $(m,j)$,
such that $4Gm:=\m$ and $4Gj:=\m\ell\sqrt{1-\eta^2}$.

Choosing the triad field as
\bsubeq\lab{5.2}
\be
b^0=Ndt\,,\qquad b^1=F^{-1}dr\,,\qquad b^2=r(d\vphi+N_\vphi dt)\, ,
\ee
the Riemannian connection takes the form
\be
\om^{01}=-\a b^0+\b b^2\, ,\qquad \om^{02}=\b b^1\, ,
\qquad \om^{12}=-\b b^0+\g b^2\, ,
\ee
\esubeq
where $\a:={FN'}/N$, $\b:={rFN'_{\vphi}}/{2N}$ and $\g=F/r.$ These
objects define the Riemannian geometry of the rotating OTT black
hole in the context of PGT.

Now, based on theorem \ttwo\ from section 2, we know that the rotating OTT
black hole, being an exact solution of the BHT gravity, is also a solution
of PGT provided its Cotton tensor vanishes. Technically, the proof that
$C^{ij}=0$ is not quite simple due to the complicated structure of the
metric functions $N,F$ and $N_\vphi$. However, relying on the standard
computer algebra systems, one easily finds that $C^{ij}$ indeed vanishes.

\subsection{Asymptotic conditions and conserved charges}

A direct inspection of the rotating black hole geometry \eq{5.2} shows
that it belongs to the same class of asymptotic states as described by
Eqs. \eq{4.1} and \eq{4.3}. Hence, the results for (i) the boundary term
\eq{4.8}, and (ii) the classical central charges \eq{4.14}, remain valid
also in the rotating black hole case.

Applying formulas \eq{4.10} to the rotating OTT geometry \eq{5.2}
yields the following conserved charges:
\bsubeq\lab{5.3}
\bea
&&E=\frac{1}{4G}\left(\m+\frac{1}{4}b^2\ell^2\right)\, ,         \\
&&J=\ell\sqrt{1-\eta^2}\,E\,.
\eea
\esubeq
For $\eta=1$ the angular momentum vanishes, whereas for $b=0$ we have the
BTZ black hole with $E=m$ and $J=j$; its energy is twice as big as in GR.

\subsection{The first law of black hole thermodynamics}

The entropy for the rotating OTT black hole can be calculated in the
same manner as for the static one. Using the above expressions for $E$,
$J$, and the central charges $c^\pm=3\ell/G$, the Cardy formula yields
\be
S=2\pi\ell\sqrt{\frac{(1+\eta)E}{2G}}\,.
\ee
The Hawking temperature and the angular velocity at the outer horizon are:
\bea
&&T=\frac{1}{4\pi}\frac{\pd_r N^2}{A}\Big|_{r=r_+}
   =\frac{1}{\pi\ell}\sqrt{\frac{2\eta^2}{1+\eta}}\sqrt{GE}\,,\nn\\
&&\Om_+=N_{\vphi}\big|_{r=r_+}=\frac{1}{\ell}\sqrt{\frac{1-\eta}{1+\eta}}\,.
\eea
Then, the first law of black hole thermodynamics is automatically satisfied:
\be
T\d S=\d E-\Om_+\d J\,.
\ee

\section{Discussion and conclusions}\label{sec6}
\setcounter{equation}{0}

The OTT black hole energy was calculated already in the original paper
\cite{x12}, based on the Deser--Tekin approach \cite{x26}. Since the
Deser--Tekin formula (37) in \cite{x12} does not contain the asymptotic
terms produced by the parameter $b$, the resulting energy $E_{\rm
DT}=\m/4G$ does not depend on $b$. This result is evidently not compatible
with the first law of black hole thermodynamics. Then, Giribet et al.
\cite{x14} found certain arguments, based on interpreting $b$ as a `hair'
parameter, to transform $E_{\rm DT}$ into $E=(\m+\ell^2 b^2/4)/(4G)$, the
expression that is fully compatible with the first law \cite{x25}.

In the next paper, Giribet and Leston \cite{x15} tried to find more
convincing arguments to derive the above form of $E$. Their approach was
based on the work of Hohm and Tonni \cite{x27}, who developed a
generalized Brown--York approach to the generic form of the BHT gravity.
By restricting their considerations to the special value of $m^2$, where
the OTT black hole is admitted as an exact solution, the authors of
\cite{x15} succeeded to derive the above result for $E$, but only for the
rotating black hole, where certain ambiguity in the derivation disappears.
By improving the construction, Kwon et al. \cite{x16} obtained the
conserved charges for both the static and the rotating OTT black hole. Our
expressions \eq{5.3} for the conserved charges confirm their final
results, given in Eq. (44).

In the approach initiated by Regge and Teitelboim \cite{x17}, the
gravitational conserved charges and the improved canonical generators are
closely related to each other. An important progress in understanding
essential aspects of this relation has been achieved in the first order
approach, which allows one to find a covariant boundary term and identify
its value as a conserved charge; for an early version of the formalism,
see Nester \cite{x24}, and for a comprehensive exposition of this
approach, see Chen et al. \cite{x11}. The covariant approach has been
widely used in 4D gauge theories of gravity with a great success
\cite{x4,x11}. Moreover, it was also confirmed on a set of selected 3D
solutions \cite{x28}. Now, in order to properly understand our results in
the context of this approach, we start from a particular choice of the
covariant boundary expression (integrand) defined by the \emph{upper line}
in Eq. (234) of \cite{x11}:
\be
B_{\rm ul}(\xi):=(\xi\inn b^i)\D\t_i+\D b^i(\xi\inn\t_i)
   +(\xi\inn\om^i)\D\r_i+\D\om^i(\xi\inn\r_i)\, .               \lab{6.1}
\ee
Here, $\D X=X-\bar X$ is a difference between a field $X$ and its boundary
value $\bar X$, and $\xi$ is asymptotically a Killing vector field. The
lower line is obtained by replacing the variables $(b^i,\t_i,\om^i,\r_i)$
with their boundary values. One can verify that formula \eq{6.1}, taken in
the Riemannian limit, is not compatible with our result \eq{4.8}. This is
in fact true for all sixteen versions of $B(\xi)$, obtained from Eq. (234)
of \cite{x11} by choosing \emph{either upper or lower term} in each of the
four curly brackets separately. However, the situation is changed by
generalizing the construction of $B(\xi)$ in a way proposed by So
\cite{x18}. According to his prescription, the original Hamiltonian
boundary term $B(\xi)$ is modified by replacing each curly bracket by
a \emph{linear combination} of its upper and lower term. Applying this
prescription to Eq. (234) of \cite{x11}, one finds that its Riemannian
reduction takes the form
\be
\tB(\xi;c_3,c_4):=\xi\inn\left[c_3\om^i+(1-c_3)\bar\om^i\right]\wedge \D\r_i
        +\D\om^i\wedge\xi\inn\left[c_4\r_i+(1-c_4)\bar\r_i\right]\,,\lab{6.2}
\ee
where $c_3$ and $c_4$ are real parameters. For the particular choice
$(c_3,c_4)=(1/2,1/2)$, we have
\be
\tB(\xi;1/2,1/2):=\xi\inn\left[\om^i-\frac{1}{2}\D\om^i\right]\wedge \D\r_i
        +\D\om^i\wedge\xi\inn\left[\r_i-\frac{1}{2}\D\r_i\right]\,,\lab{6.3}
\ee
A comparison with Eqs. \eq{4.10} shows that the boundary term
$\int_{\pd\S}\tB(\xi;1/2,1/2)$ exactly coincides with our expression
$\G(\xi)$, Eq. \eq{4.8}.

Clearly, the result \eq{6.3} represents only a Riemannian reduction of a
more general So-like formula for the boundary term. With an obvious
extension of notation, this more general formula can be represented in the
form
\be
B(\xi)=B(\xi;c_1,c_2,1/2,1/2)\,.
\ee
Additional information on the general structure of $B$ can be found in
Ref. \cite{x27}, where the conserved charges of several 3D solutions were
calculated. However, the results are not sufficiently sensitive to clearly
recognize the general structure of a ``good" expression for the boundary
term in PGT, in 3D. Further work in this direction is needed.

\medskip
In conclusion, we summarize our results as follows.

(a) First, we found general criteria that allow us to study
conformally flat Riemannian spacetime configurations as solutions of PGT.
These criteria are used to show that the OTT black hole, a solution of the
BHT gravity, is a Riemannian solution of PGT.

(b) Then, we constructed a natural set of the asymptotic conditions and
calculated the conserved charges of the OTT black hole as the values of
the Hamiltonian boundary term. The expressions for the conserved charges
coincide with those found by of Kwon et al. \cite{x24} in the generalized
Brown--York approach.

(c) Finally, the obtained results are verified by showing that: (i) the
conserved charges are exactly compatible with the first law of black hole
thermodynamics, and (ii) our boundary term is in agreement with the
generalized covariant formula proposed by So \cite{x18}. On the other
hand, the OTT black hole appears to be an interesting physical example for
the generalized covariant formula.

\section*{Acknowledgements}

One of us (M.B.) would like to thank James Nester for a very instructive
discussion on the form of covariant boundary terms. This work was
supported by the Serbian Science Foundation under Grant No. 171031. The
results are checked using the computer algebra systems \emph{Reduce} and
\emph{Mathematica}.

\appendix

\section{Useful asymptotic relations}
\setcounter{equation}{0}

In the Riemannian sector of PGT, the condition $T^i=0$, calculated on the
asymptotic configurations \eq{4.1} and \eq{4.3}, leads to an additional
set of asymptotic requirements:
\bea
&&\frac{r^2}{\ell^2}B^1{_r}-\ell\Om^2{_t}=\cO_1\, ,\qquad
  B^1{_r}-\frac{\ell^2}{r^2}\Om^0{_\vphi}=\cO_1\, ,             \nn\\
&&\frac{r^2}\ell\Om^1{_r}+\Om^2{_\vphi}=\cO_1\, ,\qquad
  \frac{r^2}{\ell^2}\Om^1{_r}+\Om^0{_t}=\cO_1\, ,               \nn\\
&&\frac{B^0{_\vphi}}{\ell}+\Om^2{_\vphi}+B^2{_t}+\ell\Om^0{_t}=\cO_1\,.\lab{A.1}
\eea
Then, relying on the asymptotic form of the Schouten tensor $L_{ij}$,
\bea
&&L_{00}=\frac{1}{2\ell^2}-\frac{1}{r\ell}\left(B^0{_t}
                     +\frac{r^2}{\ell^2}B^1{_r}\right)+\cO_2\, ,\nn\\
&&L_{11}=-\frac{1}{2\ell^2}+\cO_2\, ,                           \nn\\
&&L_{22}=-\frac{1}{2\ell^2}+\frac{1}{r\ell}
  \left(\frac{r^2}{\ell^2}B^1{_r}
         +\frac{1}{\ell}B^2{_\vphi}\right)+\cO_2\, ,            \nn\\
&&L_{02}=-\frac{1}{\ell^2 r}B^0{_\vphi}
         +\frac{r}{\ell^2}\Om^1{_r}+\cO_2\, ,                   \nn
\eea
one obtains the asymptotic relations
\bea
&&\D H_{ijt}=-4a_0\Om_{ij\vphi}+\cO\left(\frac{\Om_{ij\vphi}}r\right)\,,\nn\\
&&\D H_{ij\vphi}=-4a_0\ell^2\Om_{ijt}+\cO\left(\frac{\Om_{ijt}}r\right)\,.
\eea

\section{Consistency of the boundary term}
\setcounter{equation}{0}

In this appendix, we prove the consistency of the Hamiltonian boundary
term \eq{4.8} by showing that it is a finite expression that satisfies the
variational equations \eq{4.7a}.
Using the expressions \eq{4.10} for $\cE$ and $\cJ$, as well as the
results of Appendix A, we have:
\bsubeq
\bea
\cE&=&4a_0\frac{r}\ell\left(\Om^0{_\vphi}-\frac{r^2}{\ell^2}B^1{_r}\right)
   +\cO_0=\cO_0\, ,                                             \\
\cJ&=&2a_0\om^i{_\vphi}b_{i\vphi}
      +4a_0\ell^2L_{ij}\om^i{_\vphi}b^j{_\vphi}                 \nn\\
   &=&-4a_0r\left(\frac{B^0{_\vphi}}\ell+\Om^2{_\vphi}\right)
                                    -4a_0\ell r^2L_{02}+\cO_0\,,\nn\\
   &=&-4a_0r\left(\Om^2{_\vphi}+\frac{r^2}\ell\Om^1{_r}\right)
                            +\cO_0=\cO_0\, ,
\eea
\esubeq
which completes the proof of finiteness.

In a similar manner,
\bea
\d\cE&=&\frac{1}{2}\left(\om^{ij}{_t}\d H_{ij\vphi}
                          +\d\om_{ij\vphi}H^{ij}{_t}\right)   \nn\\
&&+\frac14\left(\D H_{ij\vphi}\d\om^{ij}{}_t-\D\om^{ij}{_t}\d H_{ij\vphi}
  -\D H_{ijt}\d\om^{ij}{}_\vphi+\D\om^{ij}{_\vphi}\d H_{ijt}\right)\,,\nn\\
&&=\frac{1}{2}\left(\om^{ij}{_t}\d H_{ij\vphi}
                          +\d\om_{ij\vphi}H^{ij}{_t}\right)+\cO_1\, ,
\eea
whereas the proof for $\d\cJ$ is trivial. Thus, the variational equation
\eq{4.7a} is satisfied.

\end{document}